\documentclass[fleqn,twoside]{article}
\usepackage[cp1251]{inputenc}
\usepackage{multicol,ifthen}
\usepackage[english]{babel}
\usepackage{amssymb,amsmath,amsfonts,graphicx}
\usepackage{arxiv}

\nazva{Metastable Quasimolecules in Excited Gases}

\font\Bbb=msbm10 scaled \magstep1 \font\Bbbscr=msbm7 scaled
\newfam\Bbbfam
\textfont\Bbbfam=\Bbb \scriptfont\Bbbfam=\Bbbscr
\scriptscriptfont\Bbbfam=\Bbbscrscr
\def\Bbb{\Bbbfam}

\font\Cal=msbm10 scaled \magstep1 \font\Calscr=msbm7 scaled
\newfam\Calfam
\textfont\Calfam=\Cal \scriptfont\Calfam=\Calscr
\scriptscriptfont\Calfam=\Calscrscr
\def\Cal{\fam\Calfam}
\def\beq{\begin{equation}}
\def\eeq{\end{equation}}

\def\gappeq{\mathrel{\rlap {\raise.5ex\hbox{$>$}}
{\lower.5ex\hbox{$\sim$}}}}

\def\lappeq{\mathrel{\rlap{\raise.5ex\hbox{$<$}}
{\lower.5ex\hbox{$\sim$}}}}

\newcommand{\lsim}{\raise.3ex\hbox{$<$\kern-.75em\lower1ex\hbox{$\sim$}}}
\newcommand{\gsim}{\raise.3ex\hbox{$>$\kern-.75em\lower1ex\hbox{$\sim$}}}
\newcommand{\eq}[1]{\begin{equation} #1 \end{equation}}
\newcommand{\ml}[1]{\begin{multline} #1 \end{multline}}
\udk{31.10.+s, 31.50.+w}
\nazvacol{Metastable Quasimolecules  }%
\avtor{V. N. MALNEV, R. A. NARYSHKIN}%
\avtorcol{V. N. MALNEV, R. A. NARYSHKIN}%

\inst{Taras Shevchenko Kiev National University, Physics Department}
\adr{(6 Glushkova Prosp., 03022 Kiev, Ukraine\\ E-mail: malnev@i.com.ua, naryshkin@univ.kiev.ua)}

\begin{document}
\setcounter{page}{001}%

\maketitl

\begin{multicols}{2}

\anot{Quasimolecules, which consist of two differently excited atoms
in a resonantly excited gas, are considered. The energy of
dissociation and typical sizes of such molecules are calculated in
the first order of quantum-mechanical perturbation theory with
the help of the dipole-dipole interaction operator. It is shown that
there exist metastable quasimolecules, whose dipole radiative transition
to the ground state (two non-excited atoms) is forbidden.
The lifetime of such molecules is estimated and it is shown that quasimolecules
may considerably affect the transport processes in a resonantly excited
gas. }

\section{Introduction}

It is well known (see e.g. \cite{LL3}) that the interaction energy between atoms
which do not form stable molecules can be calculated by means of
quantum-mechanical perturbation theory using a dipole-dipole
interaction operator. Particularly, the interaction between atoms in
the ground state is described with the help of the van-der-Waals
potential
$$U_{VW}(R)=-\frac{\mathrm{const}}{R^6},\quad R\gg a,$$
where $R$ is a distance between the atoms, and $a$ is a typical size
of an atom. The potential $U_{VW}(R)$ is obtained in the second
order of perturbation theory and always corresponds to
attraction. The above-mentioned forces are the so-called van-der-Waals
forces, which form a relatively shallow potential well (approximately
$10^{-3}$ eV for $\mathrm{He}_2$ \cite{Buck} and $2\times 10^{-2}$ eV
for $\mathrm{Kr}_2$ \cite{Tanaka} that corresponds to approximate 
temperatures 10 K and 200 K respectively), so that the formation of
stable molecules in such gases at room temperature does not happen.

The situation changes cardinally if one of the atoms is in an
excited state, from which the dipole transition to the ground state is
possible. In this case, the dipole-dipole interaction operator gives
a non-zero contribution even in the first order of perturbation
theory due to the fact that non-perturbed wave functions must account for 
an exchange of excitations between the atoms. The
transition of the excitation from one atom to the other in this case
replaces the exchange interaction that leads to the chemical
valence. Such an interaction between the atoms is conventionally
called a resonance interaction. It is described with the help of
the potential
$$U_R(R)=\pm \frac{\mathrm{const}}{R^3},$$ and can have a character of not
only attraction, but also repulsion.

The dissociation energies of bound states, which can be formed at
the expense of the resonance dipole-dipole interaction, turn out to be of the order of $1$
eV according to the below-presented estimations. This means that quasimolecules, which consist of atoms in the
ground and an excited states, can be formed at room temperature. It
should be emphasized that a relatively large energy of the resonance
interaction is related, as mentioned above, to the exchange of
excitations realized between identical atoms through the
dipole-dipole interaction operator.

It is worth noting that in the case of differently excited identical
atoms, when the dipole transitions from excited states to the ground
state are forbidden, the formation of quasimolecules (so-called
excimers) is also possible \cite{excsimer}. In contrast to resonantly
excited quasimolecules, when the attractive part of the potential
energy of the interaction as a function of $R$ can be calculated
analytically, the potential energy curves of the molecular states of
excimer molecules, as known, can be calculated only by numerical methods.

The presence of even relatively low concentrations of resonantly
excited molecules may essentially affect the collective properties
of excited gases and plasma. The thermodynamic functions of resonantly excited gases 
were first considered in \cite{malnev}.
Quasimolecules of two differently excited one-electron atoms in such
gases were discussed in \cite{vdovin}, and a role of
quasimolecules in resonantly excited gases was considered in
\cite{naryshkin}.

The purpose of the present paper is an analytical calculation of the
long-range part of the potential energy curves of molecular states
of the resonantly excited quasimolecules formed from identical atoms with the
closed electron shells and atoms with one valent electron that do
not form stable molecules in the ground state at room temperature.
We also show that a radiative transition of resonantly excited
quasimolecules to the ground state (two non-excited atoms) is
forbidden in the dipole approximation. Therefore, the considered
quasimolecule is metastable and its lifetime appears to be of the
order of $10^{-5}$ sec.

\section{Quasimolecule of Helium He$_2^*$}
\label{sec}

Consider an interaction of two helium atoms, one of which is in the
ground state, while the other is in the excited $P-$state. Let us
denote the nuclei of these atoms as A and B and assume that the distance
between them equals $R.$ We will use two coordinate systems,
which describe the electrons of the atoms, and match their origins
with the corresponding nuclei (A and B). Next, we direct the $z$ axes of
these coordinate systems along the line that connects these
nuclei, and denote the electrons belonging to the first
atom (nucleus A) by indices 1 and 2. The electrons belonging 
to the second atom (nucleus B) are denoted by indices $1'$ and $2'.$

The electron wave function of atom A (in the ground state)
calculated with the help of the variational method looks like
\cite{Flugge} \eq{\Psi(1,2)=\frac{\alpha^3}{\pi a^3}\;e^{-\,\alpha
\textstyle \frac{r_1+r_2}{a}},} and represents
the product of two hydrogen-like wave function for each electron in
the field of the nucleus with an effective charge $Z_\text{eff}=\alpha=27/16:$
\eq{\label{psi0}\widetilde{\psi}_0(r)\equiv|\widetilde{0}\rangle=\frac{\alpha^{3/2}}{\sqrt{\pi}\,
a^{3/2}}\,e^{-{ \;\alpha \textstyle \frac{r}{a}}}.} Here and below,
$a$ designates Bohr's radius: \eq{a=\frac{\hbar^2}{m_ee^2}\simeq
0.529 \text{\AA}.}

We take the electron wave function of the excited $P$-state of a
helium atom in the form of an antisymmetrized product of the hydrogen-like
wave functions of an $S$-electron in the field of a nucleus with charge
$Z=2$ and a $P$-electron in the field of the atomic residue with
charge $Z=1$ \cite{Flugge}:
\eq{\Psi(1',2')=\frac{1}{\sqrt2}\left[\psi_0(1')\varphi_i(2')\pm\psi_0(2')\varphi_i(1')\right],\quad
i=x,y,z,} where
\eq{\label{psi00}\quad{\psi}_0(r)\equiv|0\rangle=\frac{\beta^{3/2}}{\sqrt{\pi}\,
a^{3/2}}\,e^{- \beta \textstyle \frac{r}{a}},\quad \beta=2,}
 and
\eq{\label{5}{\left\{\begin{array}{l}\varphi_{x}(\mathbf{r})\equiv|x\rangle=
R_{n1}(r)\,\sqrt{\frac{3}{4\pi}}\;\sin{\theta}\;\cos{\varphi},
\\\varphi_y(\mathbf{r})\equiv|y\rangle=
R_{n1}(r)\,\sqrt{\frac{3}{4\pi}}\;\sin{\theta}\;\sin{\varphi},\\
\varphi_z(\mathbf{r})\equiv|z\rangle=R_{n1}(r)\,\sqrt{\frac{3}{4\pi}}\;\cos{\theta}.\end{array}\right.}}
The degenerated wave functions (\ref{5}) are chosen such
that they transform as vector components under rotations of the
coordinate system.

Henceforth, we restrict ourselves by the description of the
interaction of atoms in the dipole-dipole approximation. As is well
known \cite{malnev}, the resonance interaction (between atoms in
the ground and an excited states) is inversely proportional to $R^3$
and shown to be the most intensive for low-excited states.
Therefore, we will consider the lowest excited $P-$state with the
principal quantum number $n=2$, for which
\eq{R_{21}(r)=\frac{2}{\sqrt3}\;\frac{\gamma^{5/2}}{a^{3/2}}\;\frac{r}{a}\;e^{-\gamma\textstyle
\frac{r}{a}},\quad \gamma=\frac12.}
The wave function in the zero-order approximation is a linear combination
\eq{\Psi(1,2,1',2')=\sum_{i=1}^{12}C_i \psi_i(121'2')} of the
following 12 wave functions:
\eq{\label{funx}{|\widetilde{0}\widetilde{0}0x\rangle,|\widetilde{0}\widetilde{0}x0\rangle,
|0x\widetilde{0}\widetilde{0}\rangle,|x0\widetilde{0}\widetilde{0}\rangle,}}
\eq{\label{funy}{|\widetilde{0}\widetilde{0}0y\rangle,|\widetilde{0}\widetilde{0}y0\rangle,
|0y\widetilde{0}\widetilde{0}\rangle,|y0\widetilde{0}\widetilde{0}\rangle,}}
\eq{\label{funz}{|\widetilde{0}\widetilde{0}0z\rangle,|\widetilde{0}\widetilde{0}z0\rangle,
|0z\widetilde{0}\widetilde{0}\rangle,|z0\widetilde{0}\widetilde{0}\rangle.}}
Wave functions (\ref{funx})--(\ref{funz}) are the products of
the one-particle wave functions (\ref{psi0}), (\ref{psi00}), and (\ref{5})
with the electron coordinates taken in the order $121'2',$ e.g.
$\psi_6(121'2')=|\widetilde{0}\widetilde{0}y0\rangle=\widetilde{\psi}_0(r_1)\widetilde{\psi}_0(r_2)
\varphi_y(\mathbf{r}_{1'})\psi_0(r_{2'}).$

The interaction operator in the dipole-dipole approximation has the
form
\eq{V=\frac{(\mathbf{d}_1+\mathbf{d}_2)(\mathbf{d}_{1'}+\mathbf{d}_{2'})-
3(\mathbf{d}_1+\mathbf{d}_2,\mathbf{n})(\mathbf{d}_{1'}+\mathbf{d}_{2'},\mathbf{n})}{R^3},}
where $\mathbf{n}=\mathbf{R}/R,$ and the dipole moments of electrons in
their proper coordinate systems are equal to
\eq{\mathbf{d}_{1,\,2}=-e\,\mathbf{r}_{1,\,2},\quad\mathbf{d}_{1',\,2'}=-e\,\mathbf{r}_{1',\,2'}.}

In the first order of the stationary perturbation theory the secular
equation \eq{\mathrm{det} |V_{ij}-\Delta E\,\delta_{ij}|=0}
(determinant of the 12-th order) constructed of matrix elements of the
interaction operator on functions (\ref{funx})--(\ref{funz}) brakes
into 3 blocks for each of the set of functions
(\ref{funx})--(\ref{funz}). Namely, for sets
(\ref{funx})--(\ref{funy}) we have

\eq{\label{secxy}\left|\begin{matrix}
  - \Delta E & 0 & A & A \\
  0 & -\Delta E & A & A \\
  A & A & -\Delta E & 0 \\
  A & A & 0 & -\Delta E \\
\end{matrix}\right|=0,}
and, for set (\ref{funz}), we have a similar equation with value $-2A$ in
the place of $A.$ \\
Here, constant $A$ corresponds to 24 non-zero matrix elements of the
dipole-dipole interaction operator on wave functions
(\ref{funx})--(\ref{funz}). This value can be
calculated as \ml{A=\langle\widetilde{0}\widetilde{0}x0|
V|x0\widetilde{0}\widetilde{0}\rangle=\frac{1}{R^3}\; \langle\,
0\,|\,\widetilde{0}\,\rangle^2\; \langle\, x\,|\,
d_x|\,\widetilde{0}\, \rangle^2\\=\frac{e^2}{R^3}\left[\int
dV\;\frac{(\alpha\beta)^{3/2}\;e^{-(\alpha+\beta)\textstyle
\frac{r}{a}}}{\pi\,a^3} \right]^2\\\times\left[\int dV
\;\frac{\alpha^{3/2}\,e^{-{ \;\alpha \textstyle
\frac{r}{a}}}}{\sqrt{\pi}\, a^{3/2}}\;\varphi_x(r,\theta,\varphi)\;
r \sin{\theta}\cos{\varphi}
\right]^2\\
=\frac{e^2a^2}{R^3}\;\frac{2^{16}\;\alpha^6\beta^3\gamma^5}{(\alpha+\beta)^6\,(\alpha+\gamma)^{10}}
.\phantom{sdfgddfg1111111111.}}
The substitution of constants $\alpha=27/16,$ $\beta=2,$ and
$\gamma=1/2$ gives \eq{A\simeq 0.06\,\frac{e^2a^2}{R^3}.}

From the secular equation, which reduces to 3 equations of the 4-th
order, namely to $\Delta E^4-4A^2\Delta E^2=0$ for $x,y$-functions
(\ref{funx}), (\ref{funy}) and to $\Delta E^4-16A^2\Delta E^2=0$ for
$z$-functions (\ref{funz}), we obtain the energy of states, into which the
12-fold degenerated energy level of the system of two non-interactive
atoms is split: \eq{\Delta E=\pm 2A,0,0} for $x,y$-functions and
\eq{\Delta E=\pm 4A,0,0} for $z$-functions.

The 6-fold degenerated level with zero energy and the levels with positive
energies ($+2A$, $+4A$) are ``dissociable'' levels and they do not form
any bound states. We are interested in the levels with negative
energies: the doubly degenerated (in $x$ and $y$) level with the energy
$-2A$ and the wave function (a linear combination whose coefficients are
eigenvectors of the correspondent secular equation)
\eq{\label{psix}\Psi^{(x)}_{-2A}=\frac12\left(|\widetilde{0}\widetilde{0}0x\rangle+|\widetilde{0}\widetilde{0}x0\rangle-
|0x\widetilde{0}\widetilde{0}\rangle-|x0\widetilde{0}\widetilde{0}\rangle\right),}
\eq{\label{psiy}\Psi^{(y)}_{-2A}=\frac12\left(|\widetilde{0}\widetilde{0}0y\rangle+|\widetilde{0}\widetilde{0}y0\rangle-
|0y\widetilde{0}\widetilde{0}\rangle-|y0\widetilde{0}\widetilde{0}\rangle\right),}
and the non-degenerated level with the energy $-4A$ and the wave function
\eq{\label{psiz}\Psi^{(z)}_{-4A}=\frac12\left(|\widetilde{0}\widetilde{0}0z\rangle+|\widetilde{0}\widetilde{0}z0\rangle+
|0z\widetilde{0}\widetilde{0}\rangle+|z0\widetilde{0}\widetilde{0}\rangle\right).}
The eigenvalues (energies) are written as indices in wave
functions (\ref{psix})--(\ref{psiz}).

According to the classification of electron levels of diatomic
molecules (see e.g. \cite{LL3}), the considered excited
$P$-states of a quasimolecule with energies $-2A$ and $-4A$ belong to
$^1\Pi_u$ and $^1\Sigma_g$ states, respectively.

Let us now consider determination of
the equilibrium distance $R$ between the nuclei. It is impossible to
find it by accounting for only
the long-range dipole-dipole attractive forces. Therefore, we have to consider repulsion forces that
act at small distances between the atoms. We calculate the
equilibrium distance using the following simple model: we
assume that the electron shells of two atoms are not deformed by the
interaction, and that the equilibrium distance is attained only when the maxima of
their electronic probability densities are in contact with each other 
(notice that even this configuration corresponds to the
essential overlapping of the wave functions).

The maximum of the electron probability density of a helium atom in the
ground state can be found from the equation
\eq{\frac{d}{dr}\left(r^2
\widetilde{\psi}_0^2(r)\right)=\frac{d}{dr}\left(
\frac{\alpha^3}{\pi a^3}\;r^2\;e^{-\,2 \alpha \textstyle
\frac{r}{a}}\right)=0,} which gives
\eq{\label{rmax}r_{\mathrm{max}}=a/\alpha=16/27\, a\simeq 0.6\, a.}
In a similar way, we find the maximum of the electron probability density of a
helium atom in the excited $P-$state. In this case, the most
probable value of $r$ equals \eq{r^*=4\, a,} but the wave
function, in addition, depends on the angle coordinates $\theta, \; \varphi$.
Accounting for that, we find the maximum of the distribution of the 
the probability density of $x, y$-wave functions
$|\varphi_\perp|^2=|\varphi_x|^2+|\varphi_y|^2$ to be at
\eq{\label{rmaxp}r_{\mathrm{max}}=r^*\,\sin^2{\theta}=4\,a\,\sin^2{\theta}.}
From contact condition of the surfaces (\ref{rmax}) and
(\ref{rmaxp}), we find the unknown equilibrium distance:
\eq{R^{(x,y)}_0\simeq1.203\, a,\quad (\theta\simeq 28.4^\circ).}
Similarly, for $z$-functions we have:
\eq{r_{\mathrm{max}}=r^*\,\cos^2{\theta}=4\,a\,\cos^2{\theta},}
wherefrom we get \eq{R^{(z)}_0\simeq4.6\, a,\quad (\theta=
0^\circ).}

At last, knowing the equilibrium distances between the nuclei, we
are able to calculate the corresponding binding energies (dissociation
energies with the opposite sign): \eq{\label{Exy} \Delta
E^{(x,\,y)}=-1.875\; \mathrm{eV},} \eq{\label{Ez}\Delta
E^{(z)}=-0.067\; \mathrm{eV}.}

We must emphasize that the total energy of a quasimolecule relative to
the energy of non-interacting helium atoms in the ground state is
larger than the calculated dissociation energy by the value (experimental data \cite{GG}): 
\eq{E(1s2p)-E(1s^2)\simeq
-57.787\;\mathrm{eV}+79.005 \;\mathrm{eV}\simeq 21.22\;\mathrm{eV}.}
This energy must be used in considerations of radiation energy
of quasimolecules (see the next section).

\section{Lifetime of He$^*_2$ quasimolecules}

Let us now discuss the lifetime of quasimolecules. For
such a study, we have to use the radiation theory of molecular
systems. The total probability of radiation is written in the form
of a perturbation theory series, each member of which represents a
multipole radiation of a certain type (electric or magnetic) and order
(dipole, quadrupole etc). Moreover, it is well known that the
increase of the order of multipolarity by 1 decreases the
radiation probability by $\sim (ka_0)^2$ times, where $a_0$ is a
characteristic size of the radiative system and
$k=2\pi/\lambda=\omega/c$ is a wave number of the radiated light
\cite{LL4}.

In our case of a quasimolecule in the state $^1\Pi_u,$ the value of $ka_0$ is
of the order of $10^{-2}$ ($\hbar\omega\sim 20$ eV, $a_0 \sim 1
\text{\AA}$), which gives the decrease of the radiation probability in
$10^{4}$ times when we proceed to the radiation with multipolarity by 1 order higher. 
For example, a characteristic time of the dipole
radiation of an excited atom is of the order of $10^{-9}$ sec, and in
the case where a dipole transition is forbidden, the lifetime
increases by 4 orders, i.e. up to $10^{-5}$ sec. Let us show that such
a situation indeed takes place for helium quasimolecules
He$^*_2$ in the state $^1\Pi_u.$

A dipole matrix element for the transition of a quasimolecule to
the ground state (two free atoms) is 
\eq{\label{dip}\mathbf{d}_{\mathrm{fi}}\equiv\langle
\Psi^{(x,y,z)}|\;\mathbf{d}_1+\mathbf{d}_2+\mathbf{d}_{1'}+\mathbf{d}_{2'}|
\widetilde{\psi}_0(1)\widetilde{\psi}_0(2)\widetilde{\psi}_0(1')\widetilde{\psi}_0(2')\rangle,}
where $\Psi^{(x,y,z)}$ are the correct wave functions of the zero-order
approximation (\ref{psix})--(\ref{psiz}).

It is easy to see that after the interchange of electron pairs 1,2
and $1',2'$ which describe physically identical states, wave
functions (\ref{psix}) and (\ref{psiy}) change their signs as well
as the integral (\ref{dip}). As a consequence, we get that this
integral is equal to zero \eq{\langle
\Psi^{(x,y)}|\;\mathbf{d}_1+\mathbf{d}_2+\mathbf{d}_{1'}+\mathbf{d}_{2'}|
\widetilde{\psi}_0(1)\widetilde{\psi}_0(2)\widetilde{\psi}_0(1')\widetilde{\psi}_0(2')\rangle=\mathbf{0},}
and, therefore, the dipole transition from the state $^1\Pi_u$ to the ground
state is forbidden.

However, for the $^1\Sigma_g$ state ($z$-functions (\ref{psiz})), the
symmetry does not forbid the dipole transition. In this case, only
the matrix element of the $z$-component of the dipole moment is non-zero
and turns out to be
\ml{\langle\Psi^{(z)}|\;(d_{1}+d_{2}+d_{1'}+d_{2'})_z|
\widetilde{\psi}_0(1)\widetilde{\psi}_0(2)\widetilde{\psi}_0(1')\widetilde{\psi}_0(2')\rangle\\=
-\frac{2^9\,\alpha^3\;\beta^{3/2}\;\gamma^{5/2}}{(\alpha+\beta)^3\,(\alpha+\gamma)^5}\,e\,a\simeq-0.49\,e
a.\phantom{1111111111}} The probability of a dipole radiation of the
excited system per unit time equals \cite{LL4}
\eq{w=\frac{4\omega^3}{3\hbar c^3} |\,\mathbf{d}_{\mathrm{fi}}|^2,}
wherefrom one can estimate the lifetime of the $^1\Sigma_g$ state:
\eq{\label{lifetime}\tau=\frac{1}{w}\simeq \frac{3\hbar^4
c^3}{4(\hbar \omega)^3}\frac{1}{(0.49\, e a)^2}\simeq 0.18
\;\text{ns}.} We see that the $^1\Sigma_g$ state is short-lived
and cannot a metastable one in contrast to the
$^1\Pi_u,$ whose dipole radiation is forbidden and
therefore a quasimolecule in this state is metastable and lives at
least $10^{-5}$ sec, that is essentially greater than characteristic
times of the dipole radiation $10^{-8}\div10^{-9}$ sec.

Let us note, that the given estimation (\ref{lifetime}) of
the quasimolecule lifetime agrees (by the order) with the experimental data
for short-lived excited states of helium \cite{Hill}.

\section{Quasimolecules of Atoms with One Valent Electron}

Let us consider a system, which consists of two identical atoms, and
where each atom has only one outer electron while all the other
electrons form entirely occupied electronic shell (chemical
elements of the first group). We assume that one of the atoms is in
the ground $S$-state (i.e. the valent electron has
$ns^1$-configuration), but the other one is in the excited $P$-state
($np^1$-configuration). The total electron wave function of each atom
can be represented in the form of the product of the electron wave
function $\Psi(\mathbf{r}_1,\dots,\mathbf{r}_{N})$ of entirely occupied electronic shell
and the wave function $\psi_{s,p}(\mathbf{r})$ of the outer electron:
\eq{\Psi=\Psi(\mathbf{r}_1,\dots,\mathbf{r}_{N})\psi_{s,p}(\mathbf{r}),}
where $N=2, 8, 10, 18, \dots$ is the number of electrons in the
entirely occupied electronic shells.

The entirely occupied electronic shells have zero total spin and
zero total angular momentum. Therefore, they do not take part in the
chemical combinations and the chemical valence is conditioned only
by the outer electrons \cite{LL3}. Hence, the wave function
$\Psi(\mathbf{r}_1,\dots,\mathbf{r}_{N})$ appears invariantly in all
the equations where chemical combinations are considered and
practically drops out from calculations. Taking this fact into
account, we will further use only the wave function of the valent electron
$\psi_{s,p}(\mathbf{r}),$ but will keep in mind that the entirely
occupied electronic shells must be considered for obtaining
equilibrium distances between the nuclei of the atoms.

Thus, the problem reduces to the interaction of two hydrogen-like
atoms, where one electron moves in the field of an atomic residue
with effective charge $Z(r),$ which can be calculated if the
wave function $\Psi(\mathbf{r}_1,\dots,\mathbf{r}_{N})$ is known.
But such a procedure can be done only by numerical
methods.

The angular part of the wave function of the outer electron is the
spherical function $Y_{lm}(\theta,\varphi),$ but the radial one
$R_{nl}(r)$ is considered unknown. Wave functions of the zero-order
approximation of two non-interactive atoms with account for the
excitation exchange are constructed similarly to what was done in Section \ref{sec} Namely, they are \eq{\label{wf}|0x\rangle,\quad
|x0\rangle,\quad |0y\rangle,\quad |y0\rangle,\quad|0z\rangle,\quad
|z0\rangle,} where
\eq{\psi_s(r)\equiv|0\rangle=R_{n0}(r)\;\frac{1}{\sqrt{4\pi}},}
\eq{\label{xyz}\psi_p(\mathbf{r})=
{\left\{\begin{array}{l}|x\rangle=
R_{n'1}(r)\,\sqrt{\frac{3}{4\pi}}\;\sin{\theta}\;\cos{\varphi},
\\|y\rangle=R_{n'1}(r)\,\sqrt{\frac{3}{4\pi}}\;\sin{\theta}\;\sin{\varphi},\\
|z\rangle=R_{n'1}(r)\,\sqrt{\frac{3}{4\pi}}\;\cos{\theta}.\end{array}\right.}\quad
n'\geq 2.} We emphasize that we neglect the exchange interaction of
the outer electron with the inner ones.

The dipole-dipole interaction operator is written in the form
\eq{\label{v}V=\frac{(\mathbf{d}_1\mathbf{d}_{1'})-
3(\mathbf{d}_1,\mathbf{n})(\mathbf{d}_{1'},\mathbf{n})}{R^3},} where
indices $1,1'$ correspond to electrons that belong to the
different atoms with the distance between them being equal to $R.$

After solving the secular equation constructed with the help of
interaction operator (\ref{v}) and the set of degenerate wave functions
(\ref{wf}), we get \eq{\Delta E^{(x,y)}=\pm A,\quad \Delta
E^{(z)}=\pm 2A,} where $A$ is
\eq{\label{an}A=\frac{e^2}{3R^3}\left(\int_0^\infty
r^3\,R_{n0}(r)R_{n'1}(r)\,dr\right)^2.}

In the case of a hydrogen atom, the radial functions $R_{nl}(r)$ are
known. Particularly, for $n=1$ and $n'=2,$ we have
\eq{\label{a1}A\simeq 0.555\frac{e^2a^2}{R^3}.} With the help of the
method described in Section \ref{sec}, we estimate the equilibrium
distance between the nuclei
$$R^{(x,y)}\approx 1.76\,a\quad (\theta\approx32.8^\circ),
\qquad R^{(z)}= 5\,a\quad (\theta=0^\circ),$$ and get the following
binding energies of quasimolecules \eq{E^{(x,y)}\simeq -2.77\, \mathrm{eV},\quad
E^{(z)}\simeq - 0.24\,\mathrm{eV}}  in the states $^1\Pi_u$ and
$^1\Sigma_g,$ respectively.

For other atoms, the radial functions are unknown, and we use the
experimental atomic radii $r_{\mathrm{at}}$ \cite{mendel} for
estimations of the dissociation energies of quasimolecules and assume
that $A\approx \frac{e^2a^2}{R^3}$. The results of calculations (by the
method of Section \ref{sec}) for $n=n'$ are listed
in the Table:
{\scriptsize
\begin{center}
\begin{tabular}{|c|c|c|c|c|c|c|}
  \hline
  Atom & $n$ & $r_{\mathrm{at}}$, $\text{\AA}$ &\!\!\!$R^{(x,y)}$, $a$\!\!&\!\!\!$E^{(x,y)}$,
  eV &$R^{(z)}$, $a$\!\!&\!\!\!$E^{(z)}$, eV \\
  \hline
  Li & 2 & 1.520 & 3.70 & $-$0.54 & 5.75 & $-$0.29 \\
  Na & 3 & 1.858 & 4.53 & $-$0.29 & 7.02 & $-$0.16 \\
  K & 4 & 2.272 & 5.54 & $-$0.16 & 8.59 & $-$0.09 \\
  Rb & 5 & 2.475 & 6.03 & $-$0.12 & 9.36 & $-$0.07 \\
  Cs & 6 & 2.655 & 6.47 & $-$0.10 & 10.04 & $-$0.05 \\
  \hline
\end{tabular}
\end{center}}

\bigskip

The typical sizes of quasimolecules can be roughly estimated by the
formula $R_\mathrm{M}=2R^{(z)}.$ Hence, the typical sizes of
quasimolecules from Li to Cs lie in the interval from $6$ to $10$
\AA. This means that the corresponding effective scattering
cross-sections are approximately in one order of magnitude larger than the gas-kinetic
ones.

As well as for helium quasimolecules, the dipole transition appears
to be forbidden in the case of the atoms with one valent
electron due to the fact that the symmetry of the wave function remains
unchanged with respect to the exchange of the nuclei. Thus, the quasimolecules are
metastable only in the $x,y-$states.

\section{Conclusions}

The analytic calculations of the molecular states of
diatomic quasimolecules, which are formed of the atoms in the ground
and an excited states, are carried out. The calculations are performed in
the first order of perturbation theory with the use of the
dipole-dipole interaction operator for atoms with entirely occupied
electronic shells and atoms with one valent electron. The results
describe molecular states at the distances $R$ greater than
the typical size of atoms. At distances $R^{(x,y,z)}$
(phenomenological parameters of the theory) of the order of atom
sizes, where the repulsion of atoms predominates, the potential energy is
modelled by a solid wall. The dissociation energy of such molecules, according to
our estimation, appears to be of the order of $1$ eV. In contrast to
excimer molecules, whose typical lifetime varies from $0.1$ to $10$
ns \cite{excsimer}, quasimolecules are metastable. The
estimation shows that their radiation decay is forbidden for
dipole transitions and is of the order of $10^{-5}$ sec.

The above-described quasimolecules exist in active media that are
used for the generation of laser radiation \cite{excsimer}, \cite{GG},
\cite{Hill}. It is worth noting that, even for the simplest excimer
molecule He$_2^*$ there exist over 60 molecular potential curves
\cite{excsimer}. This fact essentially complicates theoretical
analysis of systems containing a noticeable quantity of excited
atoms from the point of view of identification of certain
spectral levels.

The presence of excimers and quasimolecules 
essentially affects collective properties of the excited gases. 
For example, one
can affect the diffusion and the thermal conductivity coefficients in
excited gases with regard for the fact that quasimolecules have much
bigger effective cross sections than the gas-kinetic ones. In
particular, the presence of quasimolecules leads to a deceleration
of the processes of diffusion and thermal conductivity.

The presence of diatomic quasimolecules may also assist a more
active process of formation of clusters which consist of many atoms
\cite{lakhno}. The formation of quasimolecules with a finite
lifetime $\tau$ and their decay (dissociation) can be considered as
a photo-stimulated chemical reaction. As known \cite{LL6}, the anomalous sound absorption may be observed in
this case at the
expense of the mechanism of the second viscosity at frequencies
$\omega\tau\sim 1.$ Furthermore, a theory of
non-equilibrium phase transitions in many-particle systems with
finite lifetime is being intensively developed \cite{sugakov}. The most
essential conclusion of such a theory is that there exist additional
restrictions on the decay of systems into two coexistent
phases (in our case --- phases with different concentrations of
excited atoms), which requires additional improvements \cite{malnev2}.



\end{multicols}
\end{document}